# Electrical transport properties of graphene on SiO$_2$ with specific surface structures

K. Nagashio*, T. Yamashita, T. Nishimura, K. Kita and A. Toriumi
Department of Materials Engineering, The University of Tokyo
7-3-1, Hongo, Tokyo 113-8656, Japan
*nagashio@material.t.u-tokyo.ac.jp

**ABSTRACT** The mobility of graphene transferred on a SiO$_2$/Si substrate is limited to ~10,000 cm$^2$/Vs. Without understanding the graphene/SiO$_2$ interaction, it is difficult to improve the electrical transport properties. Although surface structures on SiO$_2$ such as silanol and siloxane groups are recognized, the relation between the surface treatment of SiO$_2$ and graphene characteristics has not yet been elucidated. This paper discusses the electrical transport properties of graphene on specific surface structures of SiO$_2$ prepared by O$_2$-plasma treatments and reoxidization.

## I. INTRODUCTION

The high field effect mobility of ~120,000 cm$^2$V$^{-1}$s$^{-1}$ has been extracted from a "suspended" graphene field effect transistor (FET) made by mechanical exfoliation of bulk graphite at 240 K.[1,2] Although practical application of graphene FET requires a reliable substrate, the mobility of graphene FET on the SiO$_2$/Si substrate is limited to ~10,000 cm$^2$V$^{-1}$s$^{-1}$, and the size of graphene is also restricted. Therefore, many researchers have looked for other substrates, such as porymethylmethacrylante (PMMA),[3] mica,[3] parylene,[4] and hexamethyldisilazane (HMDS),[5] from the viewpoint of the hydrophobicity and flatness. Recently, the drastic improvement of the mobility up to 60,000 cm$^2$/Vs has been achieved with a hexagonal boron nitride substrate.[6,7]

In terms of process robustness and electrical reliability, however, placing graphene on the SiO$_2$/Si substrate is still the main strategy. The scattering by charged impurities[8,9] has been suggested to be a main factor in the scattering centers proposed so far, such as corrugations,[10] defects,[11] and adsorbates.[12] Analysis based on the quantum scattering time has suggested that charged impurities are located within 2 nm of the graphene sheet.[13] Therefore, the key to eliminating extrinsic scattering sources is to control the SiO$_2$ surface. Two primary surface structures for silica have long been recognized:[14] a silanol group (Si-OH) with high polarization and a siloxane group (Si-O-Si) with weak polarization (Figure 1b). Although the interaction between graphene and the SiO$_2$ surface, such as the puddle formation[15] and the Dirac point shift due to the doping from the SiO$_2$ substrate,[16] has been discussed, the surface structure for SiO$_2$ has not been considered.

In this paper, we focus on the interaction between graphene and specific surface structures of SiO$_2$ produced by O$_2$-plasma treatments and reoxidization, and discuss the scattering center located at the SiO$_2$/graphene interface.

## 2. EXPERIMENTAL DETAILS

Figure 1a shows process flows for surface treatments of a SiO$_2$/p$^+$-Si wafer with an initial SiO$_2$ thickness of 100 nm. It should be noted that all SiO$_2$/Si substrates used in this study were obtained from different parts of wafers provided from the same manufacturer. It was ultrasonically cleaned by acetone and isopropyl alcohol. The SiO$_2$ thickness was reduced to 90 nm with a HF solution and followed by a de-ionized water (DIW) rinse and N$_2$ blow. This is the typical treatment, called "HF dipping" in this paper. Generally, the surface structure is known to be a silanol group.[14] Moreover, O$_2$ plasma treatment was carried out for 10 s at 60 W. It is reported that to remove hydrocarbon contaminants on the SiO$_2$ surface by O$_2$ plasma treatment results in more silanol groups.[17,18] The flow rate of O$_2$/Ar mixture gas (1:9) was 50 cm$^3$/min. This treatment is called "O$_2$ plasma". Finally, the siloxane group surface was prepared by annealing at 1000 °C for 5 min in 100%-oxygen gas flow. The water was desorbed from two silanol groups at high temperature, which formed the siloxane group,[14] as shown by a blue dotted circle in Figure 2b. It is called "reoxidation". Noted that $SiO_2$ did not grow further because oxygen diffusion in $SiO_2$ with the thickness of 90 nm was negligible at 1000 °C.[19] Then, graphene was transferred within 1 min on various $SiO_2$/Si substrates from Kish graphite by mechanical exfoliation.

In the device fabrication by electron-beam (EB) lithography, Au marks on the $SiO_2$/Si substrate were used to adjust electrode patterns to graphene. To make the above-mentioned surface treatment effective, graphene should be transferred on the surface-treated $SiO_2$/Si substrate before the Au mark formation. Otherwise, the contamination by the resist residual blinds the difference. The graphene FET device was fabricated by conventional EB lithography and lift-off techniques. Cr/Au (10/50 nm) was selected for the contact metal because the metal-induced doping was negligible for the Cr/graphene contact.[20] The four-probe electrical measurements for the as-fabricated devices were performed in a vacuum with a source/drain bias voltage of 10 mV to remove the contact resistance.[21] In order to consider the effect of the resist residual and water on graphene, the electrical measurement was again carried out after the annealing at 300 °C for 1 hour in $H_2$/Ar mixture gas or in a vacuum.

## 3. RESULTS AND DISCUSSION
### A. $SiO_2$ surface properties

To confirm the preparation of two surface structures, static contact angles on various substrates, including Si, highly oriented pyrolytic graphite (HOPG) and surface-treated $SiO_2$ were measured from photographic images of DIW droplets with ~1.5 μL by the θ/2 method, as shown in Figure 2a. Si and graphite[22] are known to be hydrophobic. Contact angles for HF dipping and $O_2$ plasma were ~6 ° and nearly zero, respectively. The hydrophilic nature of the $SiO_2$ surface is due to the hydrogen bonding between silanol group and water molecule. The increase in silanol group density by removing hydrocarbon contaminants on $SiO_2$ by $O_2$-plasma treatments enhanced the hydrophilicity, as shown in Figure 1b. On the other hand, the contact angle for reoxidation is 42.6 °. It is hydrophobic because siloxane groups have almost no hydrogen bonding with water molecules. The contact angle of the $SiO_2$ surface obtained by the dry oxidation of bare Si was also 44.5 °. Thus, the reoxidized $SiO_2$ surface is consistent with the surface just after the Si oxidation.

The $SiO_2$ surface with silanol groups is very attractive due to its negative charges. When the $O_2$-plasma-treated $SiO_2$ substrate was kept in a laboratory atmosphere for 1 day, the contact angle increased to ~25 °, as shown in Figure 2a. This phenomenon is understood to result from the re-adsorption of hydrocarbon contaminants on the $SiO_2$ surface. Moreover, it is important to reveal the difference in contact angles between HF dipping and $O_2$ plasma. $O_2$-plasma-treated $SiO_2$ substrate was dipped again in DIW. The contact angle returned to the same level as that for HF dipping. This suggests that a non-negligible amount of hydrocarbon in DIW is attached to the $SiO_2$ surface for HF dipping. Nevertheless, the hydrocarbon contamination was low in this study, compared with the previous report (~35 °),[23] since the ultrapure DIW with a high resistivity of 18.3 MΩcm and very low organic contaminant of ~5 ppb was used with the great care in this study.

The existence of hydrocarbon on the HF-dipped $SiO_2$ surface was confirmed by X-ray photoelectron spectra, as shown in Figure 2b. When HF-dipped $SiO_2$ is treated by $O_2$ plasma, the intensity of C1s peak is reduced. However, when the $O_2$-plasma-treated $SiO_2$ is kept for 1 day, it increases considerably. The difference between HF dipping and $O_2$ plasma is suggested to be due to the hydrocarbon on the silanol groups. Based on these results, three different $SiO_2$ surface structures, (i) hydrocarbon-added silanol, (ii) hydrocarbon-free silanol, and (iii) siloxiane, were prepared, as shown in Figure 1b.

It should be emphasized that actual $SiO_2$ surfaces are not as simple as the schematic drawings shown in Figure 1b. The water and hydrocarbons will be adsorbed on the silanol surface with time. Moreover, siloxane groups are rehydrated (siloxane to silanol) because the siloxane group is stable only at high temperatures.[14] Thus, all three $SiO_2$ surfaces change with time to reduce their surface energies. Therefore, it was important that graphene was transferred within 1 min. after the surface treatments.

The topography of surface-treated $SiO_2$ was measured by an atomic force microscope (AFM) in a non-contact mode. Histograms of the corresponding height distribution over the 500 × 500 nm$^2$ regions are presented in Figure 2c. With HOPG as a reference, the sharp distribution is consistent with the apparatus limitation.[24] HF dipping is smoother than the other two SiO$_2$ surfaces. The difference in the roughness, however, was negligible in the three SiO$_2$ surfaces because the root mean squares (RMS) for HF dipping, O$_2$ plasma, and reoxidation were 0.13, 0.19, and 0.22 nm, respectively, compared with 0.003 for HOPG.

### B. Graphene/SiO$_2$ interaction

Next, the graphene/SiO$_2$ interaction is considered from the viewpoints of the size of the transferred graphene, the step height between graphene and the SiO$_2$ surface, and the change in Raman peaks of graphene. Figure 3a shows optical micrographs of graphene flakes on HF-dipped, O$_2$-plasma-treated and reoxidized SiO$_2$ substrates. Generally, graphene is 10 μm in diameter for HF dipping. However, graphene with a diameter of ~100 μm can be reproducibly obtained for O$_2$ plasma.[25] The strong interaction between the silanol group and graphene is expected to reduce the van der Waals force between the bottom layer and second layer in the thick graphite film. In other words, the transfer process can be controlled by the surface treatment.

Figure 3b shows the step height of graphene measured by AFM on three kinds of SiO$_2$ surfaces. The height of graphene on the HF-dipped SiO$_2$ substrate is highest because the water and hydrocarbons exist.[26] The relative difference in the step height is clearly observed, which supports the schematic picture of the surface structures shown in Figure 1b. Although the step heights for O$_2$ plasma and reoxidation are almost the same, the size of graphene for O$_2$ plasma is much larger than that for reoxidation. This fact indicates the stronger interaction between graphene and the O$_2$-plasma-treated SiO$_2$ surface.

Figure 3c shows the full width at the half maximum (FWHM) of the G band vs. the G band position for graphene on the three kinds of SiO$_2$ surfaces measured by the microscopic Raman spectroscopy. The wavelength, the power, and the spot size were 488 nm, ~1 mW just below the objective lens, and ~0.3 cm$^{-1}$ for a 2400 gr/mm grating, respectively. Each data point was obtained from the center position of a single graphene flake. The FWHM for HF dipping is statistically the smallest at the highest G band position, while the G band for O$_2$ plasma shows a clear red shift and larger FWHM. For reoxidation, a slight red shift of the G band took place, but the FWHM did not change significantly compared with that for HF dipping.

These red shifts are explainable from the following two reasons. One is the external doping from the substrate to graphene.[27,28] The other is the effect of the structure modulation on the lattice vibration. The shift directions for O$_2$ plasma and reoxidation are the same, even though the different charge transfer behavior is expected due to different surface structures. Therefore, red shifts of the G band are mainly considered to be related to the strength of the physical interaction between graphene and SiO$_2$. The physical origin for the structural modulation, we suspect, is the increase in strong bonding sites, as illustrated by the inset in Figure 3c. The number of bonding sites for O$_2$ plasma can exceed that for reoxidation, which causes larger FWHM, that is, the dispersion of the phonon energy for O$_2$ plasma. Therefore, the graphene/SiO$_2$ interaction is largest for O$_2$ plasma. In order to confirm it, the in-situ Raman experiment under the back-gate control is required further.

### C. Electrical transport properties of graphene

Next, the electric transport properties are discussed from the viewpoint of hysteresis, Dirac point shift, and mobility. Figure 4a shows the sheet resistivity of graphene FET devices on surface-treated SiO$_2$ substrates as a function of backgate voltage ($V_G$). All devices were measured at RT in as-fabricated condition, that is, no annealing. Gate voltage was swept from -30 to 30 V (solid lines) then back to -30V (broken lines). The peak height for O$_2$ plasma is considerably reduced from HF dipping, while the sheet resistivity curve for reoxidation is very similar to that with HF dipping. The hysteresis for O$_2$ plasma was larger than that for HF dipping. Surprisingly, there was almost no hysteresis for reoxidation. Hysteresis is often considered to be due to the orientation polarization of the water molecule.[29] There are two positions for the water molecule to adsorb on graphene. One is the top surface of graphene, and the other is the graphene/SiO$_2$ interface. If the amount of the water molecule is assumed to be the same for three kinds of SiO$_2$ surfaces, water between SiO$_2$ and graphene is considered to contribute to hysteresis. Thus, almost no hysteresis for reoxidation was achieved due to the lack of water in the graphene/SiO$_2$ interface because both the siloxane SiO$_2$ surface and graphene are hydrophobic.

Figure 4b shows the change in hysteresis for different annealing conditions at 300 °C. The hysteresis ($\Delta V_{hys}$) was defined by the voltage difference in Dirac points between the forward and reverse sweeps. Water molecules bonding to silanol groups (see (i) in Figure 1b) desorb at 220 °C, while the hydration of silanol groups (silanol to siloxane) takes place at a temperature higher than 450 °C.[30] Therefore, silanol groups for HF dipping and O$_2$ plasma exist even after annealing at 300 °C. When as-fabricated graphene FET devices were annealed in vacuum, the hysteresis was decreased for HF dipping and O$_2$ plasma. Therefore, the water was removed even from the graphene/SiO$_2$ interface. The hysteresis, however, appeared again after the exposure of devices to the laboratory atmosphere for two days. The second annealing reduced the hysteresis. The hysteresis was reversible for HF dipping and O$_2$ plasma, while there was almost no hysteresis for reoxidation even after exposure to the laboratory atmosphere for 1 month.

Figure 4c shows the direction of Dirac point shifts for post-annealing processes. Devices were annealed in H$_2$/Ar mixture gas and further in vacuum. Generally, after annealing, the Dirac point moved back to $V_G = 0$ for HF dipping and reoxidation. However, for O$_2$ plasma, the Dirac point shifted to the positive direction after both H$_2$/Ar and vacuum annealings. This strong positive shift for O$_2$ plasma suggests that negative charges exist on the SiO$_2$ surface.

Finally, mobilities for graphene FET devices for HF dipping, O$_2$ plasma and reoxidation were extracted at the carrier density of $1 \times 10^{12}$ cm$^{-2}$ using the equations $\mu = 1/en\rho$ and $n = C_{ox}(V_G-V_{Dirac})/e$, where $C_{ox}$ is the gate capacitance of SiO$_2$ and $V_{Dirac}$ is the gate voltage at the Dirac point, as shown in Figure 4d. Solid circles indicate the data before the annealing, while open squares indicate the data after the annealing. It is noted that the mobility for the device with the hysteresis is underestimated by over-counting the carrier density due to the contribution of water molecule. Although there was large variation in mobilities for HF dipping and reoxidation, the highest mobility was ~10,000 cm$^2$V$^{-1}$s$^{-1}$. However, mobilities and their variation for O$_2$ plasma are considerably low and small. These tendencies are apparent after the annealing.

To discuss the scattering center located at the SiO$_2$/graphene interface, the highest mobility for each surface treatment is selected, as indicated by red broken circles in Figure 4d. For simplicity, only the cases of no water at the graphene/SiO$_2$ interface are considered. The mobility for O$_2$ plasma degraded drastically when water was removed from the graphene/SiO$_2$ interface. In this case, the direct interaction between graphene and silanol group is inferred. Thus, it is expected that one of the main scattering centers might be negatively charged silanol groups because silanol group density has been reported as ~$5 \times 10^{14}$ cm$^{-2}$.[14,31] Therefore, the variation in the mobility was very small, and positive charges induced in graphene by negatively charged silanol groups positively shifted the Dirac point voltage for O$_2$ plasma.

However, the situation is different for HF dipping because the hydrocarbon on the silanol groups prevents the direct interaction. The Dirac point moves back to $V_G = 0$, while the variation in mobilities depends on the size of the hydrocarbon, that is, the distance between the silanol group and graphene. As shown in Figure 3b, the step height of graphene on the HF-dipped SiO$_2$ surface is higher than that on the O$_2$-plasma-treated SiO$_2$ surface, which supports this expectation.

For reoxidation, the surface is a siloxane group, which may not work as the Coulomb scattering center because the polarization is very week. Therefore, high mobility can be obtained. However, the siloxane group generally changes to silanol groups to reduce the surface energy, which causes the variation in the mobility. Based on these discussions, mobility is determined by two factors: silanol group density and the size of the hydrocarbon.

The table in Figure 5 summarizes expected values for the size of the hydrocarbon and silanol group density for various SiO$_2$ surfaces, including a HMDS-treated surface. Although these two factors are completely different for HF dipping and reoxidation, it is expected that the ratio of the two factors is similar, as shown in Figure 5. However, the ratio of two factors for O$_2$ plasma is very low. Moreover, it is reported that the mobility of graphene on HMDS could be improved because the self-assembled monolayer changes the surface structure from a silanol group to a methyl group.[5] However, organic materials, such as HMDS, are not suitable for the high temperature annealing expected during device fabrication processes. Therefore, the key approach to improving the mobility further is to optimize the siloxane SiO$_2$ surface without hysteresis due to hydrophobicity and Dirac point shift due to no doping from the siloxane group.

Recent first-principles calculations provide the binding energies of graphene on silanol and siloxane SiO$_2$ surfaces: ~13 and ~15 meV per C atom, respectively.[32] These results suggest that the interaction is very weak for both surfaces. However, in their calculations, silanol groups were structurally stabilized by bonding with the hydrogen bonding, which will reduce the interaction between graphene and silanol groups. Therefore, the interaction between graphene and silanol group seems to be underestimated in their calculation.

## 4. SUMMARY

The interaction between graphene and the SiO$_2$ surfaces with silanol and siloxane groups was studied to improve the limited mobility of graphene on the SiO$_2$ substrate. One of the main scattering centers seems to be negatively charged silanol groups with a density of $\sim 5 \times 10^{14}$ cm$^{-2}$, which degraded the mobility and shifted the Dirac point in graphene FETs on O$_2$-plasma-treated SiO$_2$ substrate. Therefore, the key approach to improving the mobility further is to optimize the siloxane SiO$_2$ surface without hysteresis due to hydrophobicity and Dirac point shift due to no doping from the siloxane group.

**Acknowledgement**

Kish graphite used in this study was kindly provided by Dr. E. Toya of Covalent Materials Co. This work was partly supported by the Japan Society for the Promotion of Science (JSPS) through its "Funding Program for World-Leading Innovative R&D on Science and Technology (FIRST Program)" and by a Grant-in-Aid for Scientific Research from the Ministry of Education, Culture, Sports, Science and Technology, Japan.

## FIGURES

**FIG. 1** (a) Process flows for three different surface treatments of the SiO$_2$/Si substrate. (b) Schematic illustration of the SiO$_2$ surface. (i) HF-dipped SiO$_2$ surface (hydrophilic), (ii) O$_2$-plasma-treated SiO$_2$ surface (hydrophilic), and (iii) reoxidized SiO$_2$ surface (hydrophobic). The covalent bonding is expressed by solid lines, while dotted lines indicate the hydrogen bonding. Three different graphene/SiO$_2$ interactions are expected.

**FIG. 2** (a) Contact angles estimated from images of DIW droplets (inset) on various substrates. The mean values for 5 sets of measurements are shown. (b) X-ray photoelectron spectra of various SiO$_2$ surfaces. The C1s peak position is shifted from its ideal position (284.2 eV) because the SiO$_2$ surface is not grounded. The lavels, "O$_2$ plasma: 10 s" and "O$_2$ plasma: 1 day", indicate the plasma-treating time and the waiting time for the air exposure, respectively. (c) Height histograms for various SiO$_2$ surfaces. The HOPG data is presented as a reference. The 500 × 500 nm$^2$ regions are measured by AFM.

**FIG. 3** (a) Optical micrographs of graphene on HF-dipped, O$_2$-plasma-treated, and reoxidized SiO$_2$ surfaces. A larger monolayer graphene can be found easily after the O$_2$-plasma treatment. (b) Step height of graphene on three different SiO$_2$ surfaces measured by AFM. Insets (top & bottom) show a typical AFM image and a schematic of the measured region, respectively. (c) G band FWHM vs. G band position of graphene on three different SiO$_2$ surfaces. Inset shows a schematic of strong bonding sites.

**FIG. 4** (a) Sheet resistivity as a function of the carrier density for graphene FET devices on surface-treated SiO$_2$/Si substrates just after the lift off. Gate voltage was swept from -30 to 30 V (solid lines) then back to -30V (broken lines). (b) Hysteresis observed during bidirectional *IV* measurements for post-annealing processes. (c) Dirac point shift for post-annealing processes. (d) Mobility extracted at the carrier density of $1\times10^{12}$ cm$^{-2}$ for graphene FET devices on surface-treated SiO$_2$/Si substrates. Solid circles indicate the data before annealing, while open squares indicate the data after annealing.

**FIG. 5** (a) Schematic illustration of the relationship between mobility (experiment) and ratio of C$_x$H$_y$ size and silanol group density (expected). (b) C$_x$H$_y$ size and silanol density for various SiO$_2$ surface and HMDS.

# Fig. 1

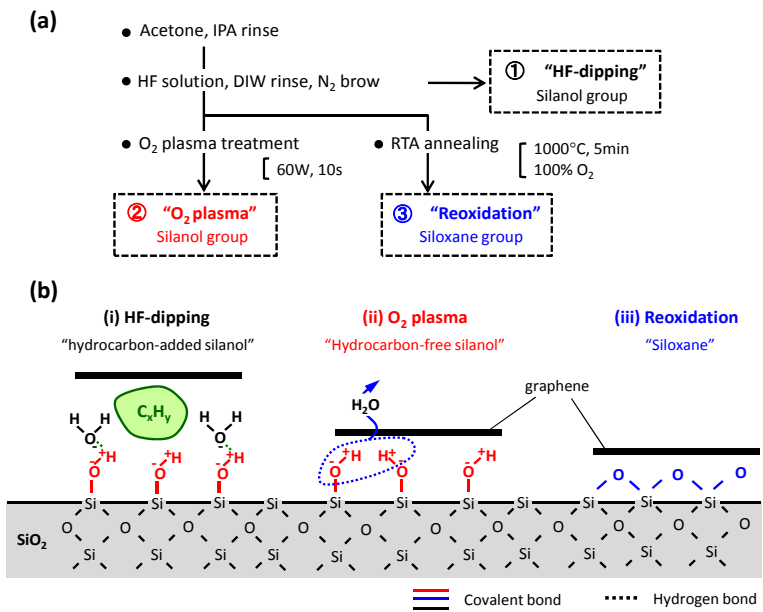

**Nagashio et al.**

# Fig. 2

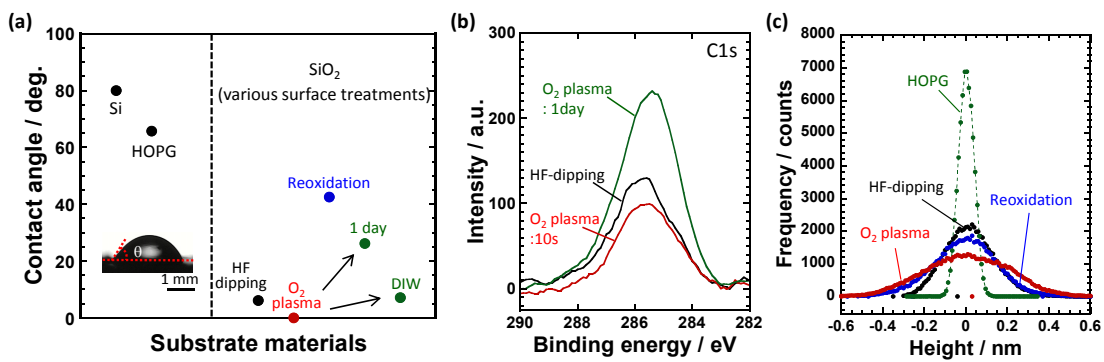

**Nagashio et al.**

**Fig. 3**

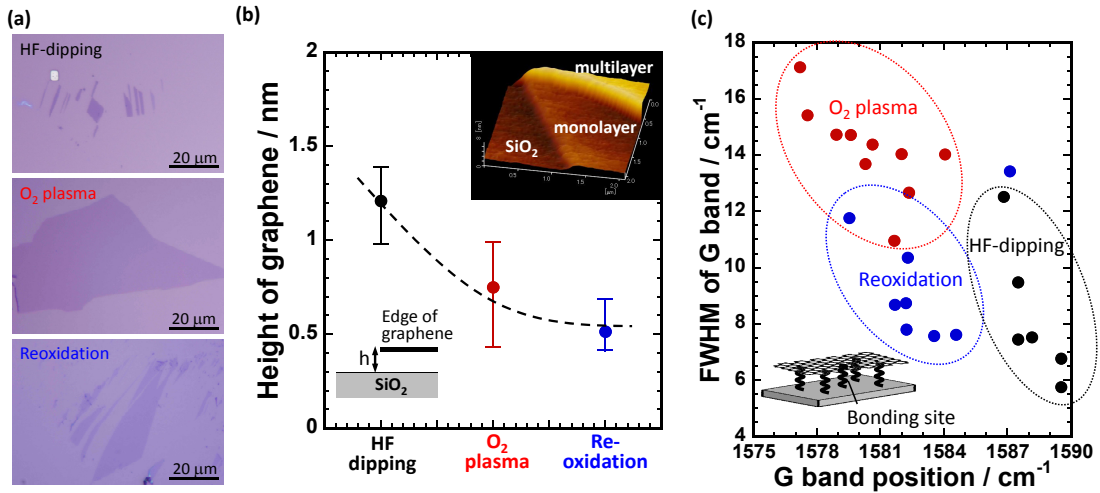

**Fig. 4**

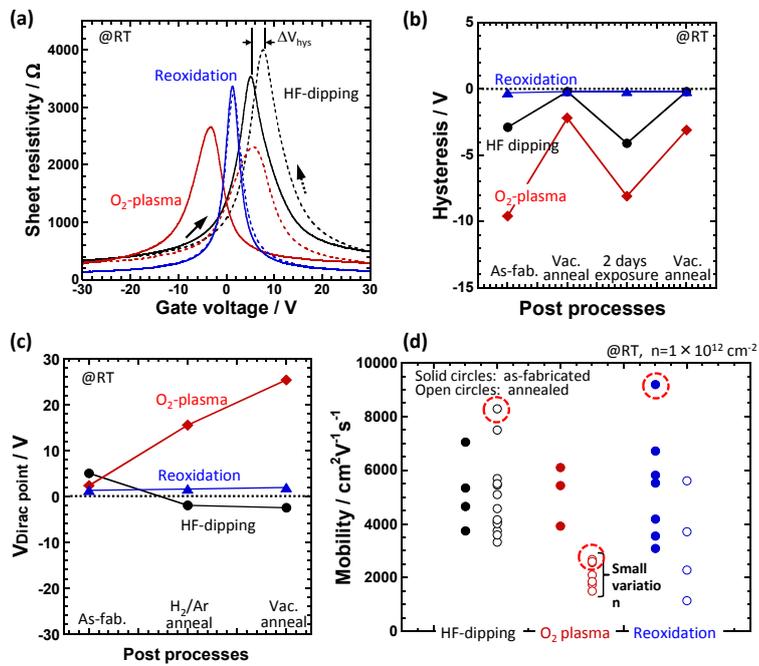

Nagashio et al.

**Fig. 5**

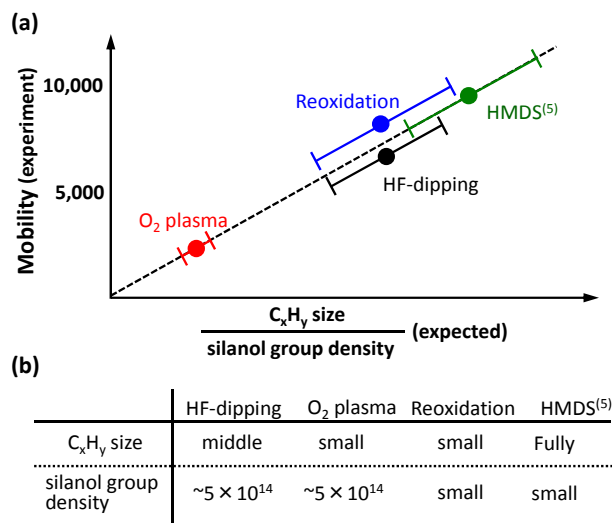

**Nagashio et al.**